\documentstyle[preprint, prl,aps]{revtex}

\begin{document} \tighten

\title{ Raman scattering by  electron-hole excitations in silver nanocrystals }
\author{H. Portales, E. Duval and L. Saviot }
\address{Laboratoire de Physicochimie des Mat\'{e}riaux Luminescents,
Universit\'{e} Lyon~1,  UMR-CNRS~5620  43, boulevard du 11 Novembre 69622
Villeurbanne Cedex, France}
\author{M. Fujii, M. Sumitomo and S. Hayashi}
\address{Department of Electrical and Electronics Engineering, Faculty of
Engineering, Kobe University, Rokkodai, Nada, Kobe 657-8501, Japan}
\date{\today}
\maketitle

\begin{abstract}
 Raman scattering experiments from silver nanocrystals embedded in films of
amorphous silica are reported. In addition to the low-frequency peak due to
vibrational quadrupolar modes, a broadband is observed in the
high-frequency range, with a maximum at about 1000 $cm^{-1}$. The linear
dependence of the position of this maximum on the inverse cluster radius is
in agreement with the Raman scattering by single or collective
electron-hole excitations.
\end{abstract}


\newpage

The spectroscopy of the metallic nanoclusters embedded in insulating
matrices is very fascinating. By Raman scattering from silver nanocrystals,
one observes in the low-frequency range (3 - 30 $cm^{-1}$), the vibrational
quadrupolar modes. The corresponding Raman peaks are relatively intense,
because of the resonance with the excitation of the dipolar electronic
plasmon \cite{Pal99}. In  the high-frequency range,  Raman scattering by
electron-hole (${\it e-h}$) excitations is expected. Some years ago  broad
high-frequency bands were observed by inelastic light scattering from
deposited $Ag$ films \cite{Mon87,Gas89,Ott92}. The bands observed in the
250 - 1000 $cm^{-1}$ range were interpreted by ${\it e-h}$ excitations in
the metallic film, their intensity and frequency being determined by the
penetration length of the polariton field in the film or by the size of
$Ag$ islands \cite{Mon87}. In other experiments, the very broad bands, that
were observed in the 1000 - 6000 $cm^{-1}$ spectral range, were attributed
to the surface enhanced Raman scattering (SERS) by the excitation of an
electron or hole from the Fermi level of silver to a level localized around
a surface defect \cite{Gas89}.

 The Raman scattering by ${\it  e-h}$ excitations was theoretically
predicted  in macroscopic metallic samples containing impurities
\cite{Fal89,Zawa90}. In this case, the ${\it  e-h}$ excitations are
overdamped. The ${\it  e-h}$ excitations localized in nanoclusters,  which
have a diameter shorter than the electronic mean free path and do not
contain impurities or defects, may be observed by Raman scattering without
being necessarily overdamped. \par

In this brief report, the high-frequency (HF) Raman broadbands, with
maximum around 1000 $cm^{-1}$, which were observed in  silica films
containing silver nanocrystals, are described. It will be demonstrated
experimentally that the frequency at the HF band maximum has an inverse
cluster radius linear dependence, and that this HF band is closely related
to the low-frequency (LF) vibrational peak. These observations will be
shown to be in agreement with the electronic Raman excitation of silver
nanocrystals.

Samples were prepared by a $rf$ cosputtering method similar to that
described in \cite{Fuj91}. Pure $SiO_{2}$ (purity 99.99\%) and $Ag$ (purity
99.99\%) were cosputtered in $Ar$ gas (purity 99.999\%) of 2.7 $Pa$ with a
$rf$ power of 200 $W$. The background pressure of the vacuum chamber was
$3\times10^{-5}\, Pa$. The substrates are $Si$ wafers. The thickness of the
films is about 400 nm. During the sputtering, the substrates were cooled by
circulating water, and the temperature was kept lower than $50^{0}$ C.
After the sputtering, the  samples were annealed in $N_{2}$ gas atmosphere
for 30 min. at $800^{0}$ C to improve the  crystallinity of the deposited
silver nanoclusters. In order to determine the size distribution, shape and
crystallinity of $Ag$ particles, the cross-sections of the samples were
observed by high resolution transmission electron microscopy. Three
different samples containing silver nanocrystals of different sizes were
studied: Sample-1 with a diameter of 3.2 nm at the maximum of the size
distribution; sample-2, diameter of 4.3 nm; sample-3, diameter of about 5
nm (as deduced from low-frequency Raman scattering).  It was verified that
the films do not  contain photoluminescent centers: By excitation in the UV
at 337.1 nm with a pulsed nitrogen laser, no luminescence was detected in
the visible range. \par

The Raman spectra were recorded with a five-grating monochromator. The
light was detected by a photomultiplier with a $GaAs$ photocathode. The
spectral response of the spectrometer-photomultiplier system was measured
by using a calibrated lamp. The 514.5 nm and 457.9 nm lines of an Argon
laser were used for the excitation. The incoming beam was at grazing
incidence and the scattered one was detected at about $\pi/2$ with respect
to the excitation. \par

In the optical absorption spectra, the maximum of the plasmon absorption
band was at about 400 nm and was extended up to more than 600 nm for all
the samples \cite{Duv00}. Figure 1 shows the Stokes-antiStokes HF reduced
Raman intensity from the film containing the biggest $Ag$-clusters, that
have a mean radius  $R\simeq 2.5$ nm, as a function of the wavenumber
$\omega (cm^{-1})$. In this figure the Raman Stokes shift is counted
positively. To obtain the reduced Raman intensity and to suppress the
dependence on temperature,  the measured Raman intensity was divided by the
thermal statistical factor $(e^{\omega/kT}-1)^{-1}$ in antiStokes, and by
$(1-e^{-\omega/kT})^{-1}$ in Stokes ($kT$ being  the thermal energy in
$cm^{-1}$). The power of the laser beam was less than 100 mW. From the
comparison between the antiStokes and the Stokes Raman intensities, the
deduced temperatures of the films under laser beam was 300 K at 457.9 nm
and 345 K at 514.5 nm. Even if the signal to noise ratio in  antiStokes is
much lower than in Stokes,  due to the thermal factor, one observes in
Figure 1  Stokes-antiStokes symmetrical spectra with  maxima at $\pm 600,
\pm 700$  $cm^{-1}$. The depolarization ratio, $I_{VH}/I_{VV}$, of the
Raman intensity for the perpendicular polarizations of excitation and
detection over the one for parallel polarizations is approximately equal to
$0.65$. \par

The  Stokes-antiStokes symmetry observed for two different excitation
wavelengths is a strong argument in favor of a Raman origin of the
broadbands seen  in Figure 1, against a possible luminescence of the
matrix, that moreover was not detected by UV excitation. As a supplementary
proof, a similar broadband appeared, with an approximately equal Raman
shift, by excitation with the 350.7 nm line of a Krypton laser. A possible
luminescence, coming from the radiative electric dipolar plasmon decay,
could be assumed. However, the decoherence time, that is due to the
coupling of the plasmon with ${\it  e-h}$ pairs \cite{Kawab66}, is very
short and close to 1 fs \cite{Ext88,Stei92}. Therefore the luminescence
transition is very unlikely. Moreover, the Stokes-antiStokes symmetry
(Figure 1) is an indication of a thermalisation, which is impossible in the
case of  luminescence, because of the too short decoherence time of the
excited plasmon state, in which this thermalisation would take place. \par

In Figure 2, the HF reduced Raman bands of the three different films are
compared with the LF peaks, which are  due to the scattering by the cluster
quadrupolar vibrational modes \cite{Pal99}. One notes that the position  of
the HF band maximum, when going from a film to another, shifts in the same
direction as the LF peak. For this comparison, it was taken into account
that the Raman scattering from the silica matrix, around $1000$ $cm^{-1}$,
is superimposed to the broadband. In the film, that contains the smallest
nanoclusters and the lowest silver concentration, the Raman lines of silica
are relatively intense and probably enhanced by the plasma oscillations.
They correspond to the different $O-Si-O$  bond stretching vibrations
\cite{Shib81}. \par

Figure 1 demonstrates that the observed HF broadband corresponds to a Raman
scattering. Its behavior is  similar to the one of the LF peak due to
cluster vibrations (Fig. 2). Furthermore, such a HF Raman band has nothing
to do with the well-known Raman spectrum of amorphous silica. Consequently,
this is necessarily a Raman band related to silver nanoclusters. The
interpretation of the HF broadband by the vibrational SERS of surface
defects, oxides or adsorbed molecules has to be rebutted. Like  in previous
experiments  \cite{Gas89,Pet83},  relatively narrow lines would be
observed.

The possible SERS by the excitation of an electron or hole from the silver
Fermi level to a level localized around a surface defect is discussed. Such
an explanation was given for the Raman broadband observed from free
deposited $Ag$ films \cite{Gas89}, with a maximum close to 3000 $cm^{-1}$.
As a matter of fact, very broad and flat bands  were observed from our
as-deposited samples, with a wavenumber at the maximum around 2500
$cm^{-1}$, that is independent of the cluster size. After annealing the HF
broadband was shifted to lower frequencies, as observed in Figure 2, its
width was narrowed, and its intensity multiplied by a factor larger than 5.
The effects of annealing on cluster crystallinity, absorption and LF Raman
scattering were analyzed in details in a previous paper \cite{Duv00}. By
transmission electron microscopy, it was observed that the crystallinity of
clusters was strongly improved. On the other hand, the width of the
absorption band was narrowed, and the LF Raman peak was shifted to
low-frequency and became more intense \cite{Duv00}. These changes by
annealing were shown to be due to the strong increase of the spatial
coherence inside clusters \cite{Duv00}. The size dependence of the
frequency at the HF band maximum, which appears after annealing (Fig. 2),
is then expected to be due to the strong improvement of the cluster
crystallinity and to be not related to defects.

To go further in the interpretation, the size dependence of the HF Raman
band is studied quantitatively in the following. The comparison of the HF
band with the LF peak is developped in order to determine the size
dependence of the HF band. In Figure 3, the wavenumber $\omega_{e}^{max}$
at the maximum of the reduced HF band is plotted against the wavenumber
$\omega_{v}^{max}$ of the reduced LF peak. The three experimental points
are aligned with the origin of coordinates. This result is very important
for the following reason. It was clearly established theoretically
\cite{Duv86}, and confirmed by experiment \cite{Pal99} that the wavenumber
$\omega_{v}$ of the LF peak is inversely proportional to the cluster radius
$R$:

\begin{eqnarray}
\label{el1}
\omega_{v}=0.85\frac{v_{t}}{2Rc}
\end{eqnarray}

\noindent
$v_{t}$ is the transversal sound velocity in silver and $c$ the vacuum
light one. In consequence, the  wavenumber $\omega_{e}$ of the HF
excitation is also inversely proportional to $R$.  From Figure 3,
$\omega_{e}/\omega_{v}\simeq 90$.

Taking into account the ratio $\omega_{e}/\omega_{v} = 90$, it was found
that the HF band shapes in the three different films can be very well
fitted by the corresponding log-normal \cite{Duv00} LF ones, if one assumes
an excitation lifetime inversely proportional to the cluster size, like
$\omega_{e}$. The ratio of the inverse lifetime (expressed in wavenumbers)
over $\omega_{e}$ is smaller than $1$ and does not change from a film to
another. This is an indication that the HF excitation is not overdamped
\cite{Zawa90}.

The inverse cluster size dependence of $\omega_{e}$ is hardly compatible
with the excitation of an electron or hole from  the  Fermi level to a
level localized around a defect inside the cluster. This shows that there
is a confinement of the excitation in the cluster. However, contrary to
internal defects, cluster surface defects can play a role  in the
desexcitation, by decreasing the ${\it  e-h}$ pair lifetime.

 The problem, which arises now, is what is the good description of ${\it
e-h}$ excitations in clusters containing more than 1000 atoms. Two
different models can be used: The shell model and the electronic plasma
modes one. In the shell model, the levels are characterized by the
principal quantum number $n$ and the angular momentum one $\ell$ (the order
of the spherical Bessel function). In this scheme, the excitation, that is
observed by Stokes Raman scattering, corresponds to the transition of an
electron from a level close to the Fermi one, to an excited level. From
group theory, the selection rules for the Raman transitions are
$\Delta\ell=0, \Delta n=\pm 1$,  or $\Delta\ell= \pm 2, \Delta n=0$. In the
simple model of electrons confined in a spherical nanocluster and
interacting with a square potential, the transition energies can be easily
calculated \cite{Kawab66,Yan92}. The energies obtained by this model have
effectively a $1/R$ dependence, but are about 9 times larger than the
observed Raman ones. However, this model is very crude and it is expected
that in ellipsoidally deformed clusters the transition energies are much
lower.

The second treatment is based on the predictions of elastodynamical
\cite{Lip89} and microcopic \cite{Nes99} models. Like the vibrations, the
electronic plasma oscillations have different multipolar spheroidal
($E\ell$) and torsional ($M\ell$) modes. The main strength of $E\ell$ modes
lies at an energy close to or higher than  the one of the surface dipolar
plasmon ($E1$) \cite{Ser89,Nest99}, and consequently cannot interpret the
Raman HF band. Among the torsional or magnetic ($M\ell$) modes, those of
the lowest energy were recently studied theoretically
\cite{Lip89,Nes99,Nes00}: The  rotational oscillations or magnetic dipolar
$M1$ modes characterized by the quantum number $\ell$ = 1, and the twist
$M2$ modes for which $\ell$ = 2. The $M2$  modes, which are odd, would be
not visible by Raman scattering.
The magnetic $M1$ modes exist when the cluster is  deformed (for example,
with a quadrupolar ellipsoidal deformation) \cite{Lip89}. They consist of
small-amplitude collective rotational  oscillations of  electrons, inside
the cluster, around an axis perpendicular to the symmetry axis of the
ellipsoid,  with the electronic oscillation amplitude vanishing at the
cluster boundary. From group theory, these $M1$ modes can be visible by
Raman scattering. Their energy, which has also the $1/R$ dependence, was
predicted to be of the order of 0.1 eV, in nanometric clusters
\cite{Lip89,Nes99}, and in consequence could be a candidate for the
interpretation of the HF Raman band. However, the experimental informations
are not sufficient, and the  theoretical studies for the electronic
torsional modes in deformed silver clusters are not yet known, so that it
is not possible to decide today what is the most relevant model to
interpret the HF Raman band.

In conclusion, it has been clearly demonstrated that the observed Raman
high-frequency broadband is in agreement with the scattering by electrons
in  silver nanoclusters. It is remarkable that the position of the
high-frequency Raman broadband is deduced from the one of the low-frequency
peak due to quadrupolar vibrational modes. This is a clear demonstration of
the inverse size dependence of the  high-frequency excitation. Such a size
dependence, that is due to  electronic confinement, is predicted in the
different  models of  ${\it e-h}$ excitations. For a better knowing of the
${\it e-h}$ excitations confined in metallic nanocrystals, new Raman
experiments are planned, and theoretical calculations of the electronic
collective mode energies in silver nanoclusters are now carried out
\cite{Nest01}.

\noindent
{\bf Acknowledgements}

E. Duval thanks warmly Dr V. O. Nesterenko for his informations on the
collective  electronic oscillations, and his relevant suggestions.

\newpage

\begin{figure}
\label{f1}
\caption{Stokes-antiStokes high-frequency Raman intensities obtained by
scattering from Sample-3. Excitations at 457.9 nm (circles), and at 514.5
nm (squares).}
\end{figure}

\begin{figure}
\label{f2}
\caption{Low- and high-frequency Raman reduced intensities.  Sample-1
(crosses), Sample-2 (triangles), and  Sample-3 (circles).  The excitation
is at 457.9 nm.}
\end{figure}

\begin{figure}
\label{f3}
\caption{Frequency, $\omega_{e}^{max}$, at the maximum of the
high-frequency Raman reduced intensity, versus the corresponding one,
$\omega_{v}^{max}$, for the low-frequency Raman intensity. The excitation
is at 457.9 nm.}
\end{figure}

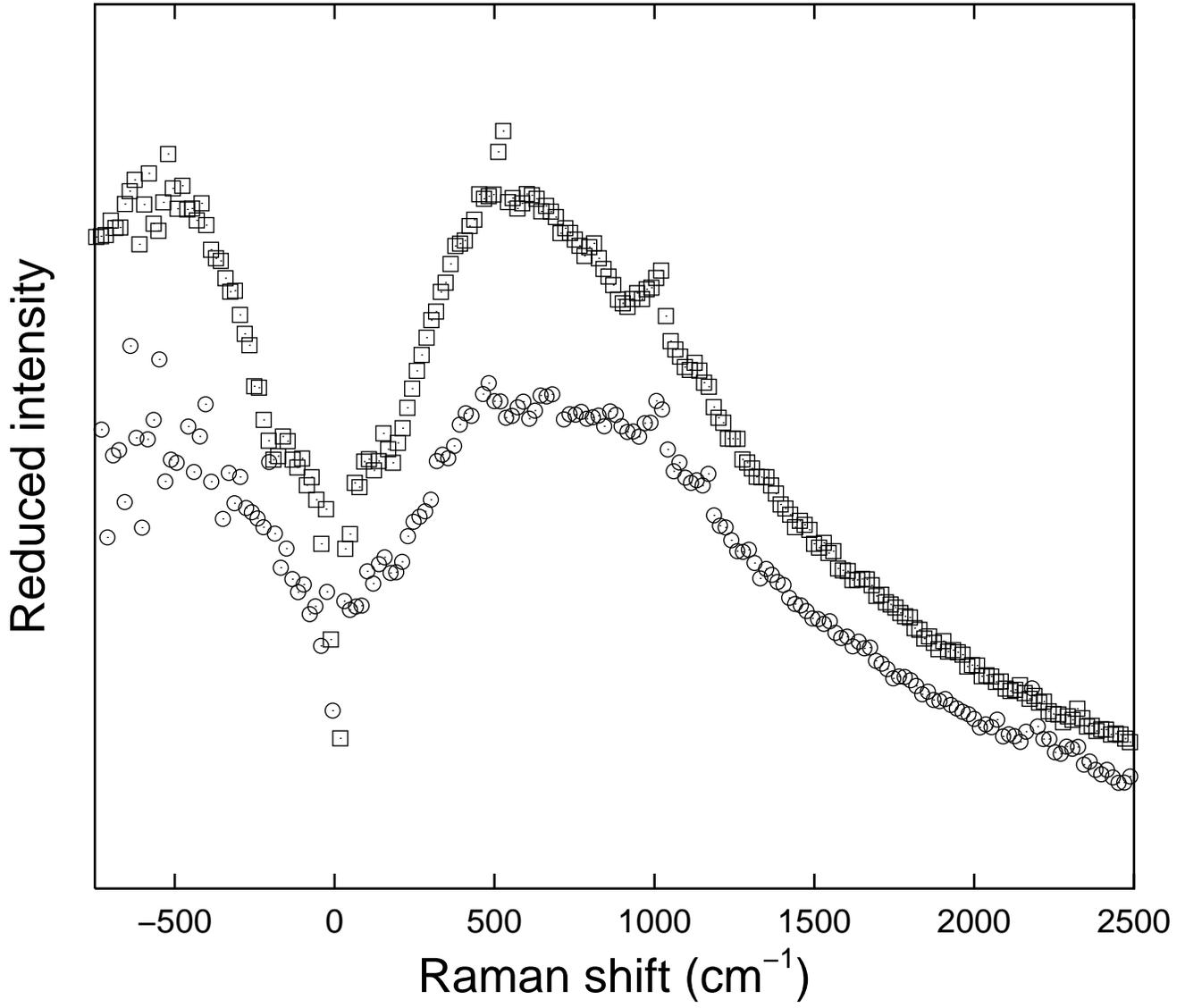
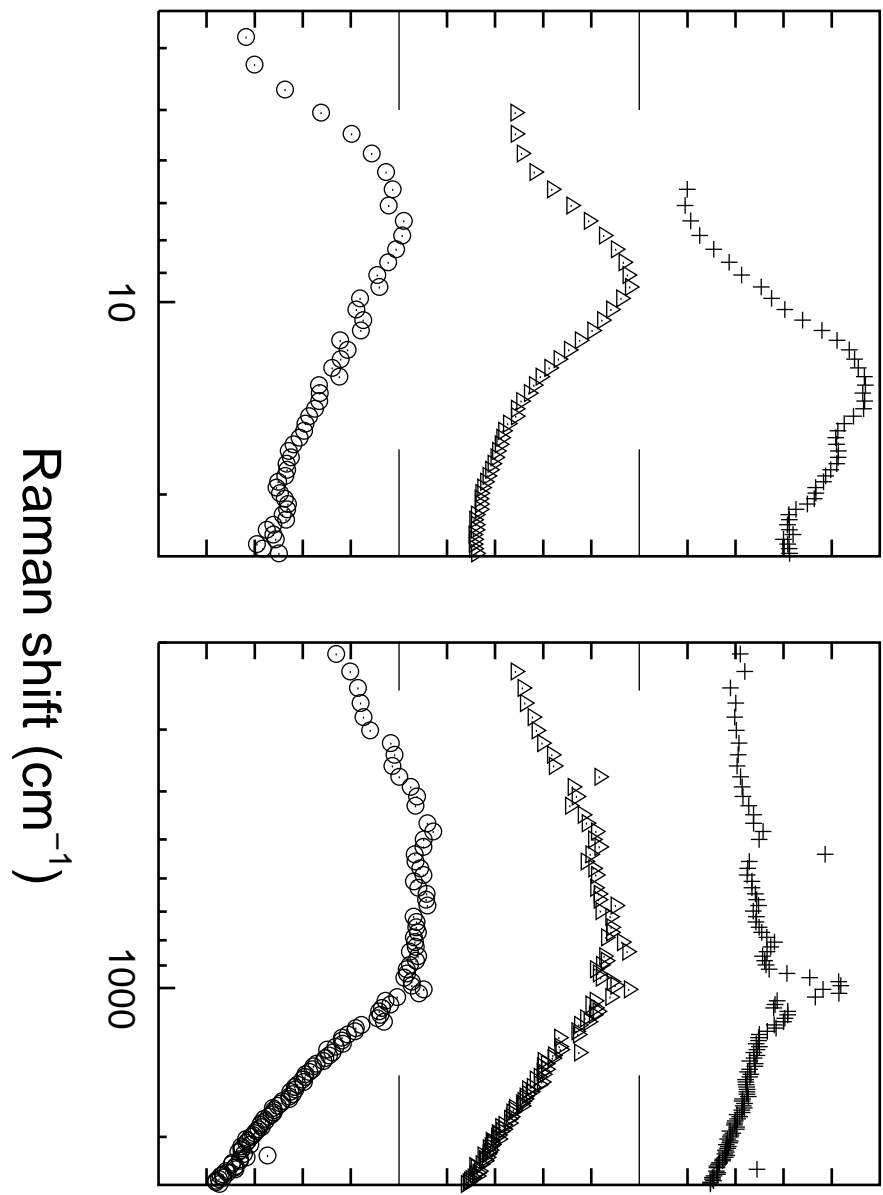
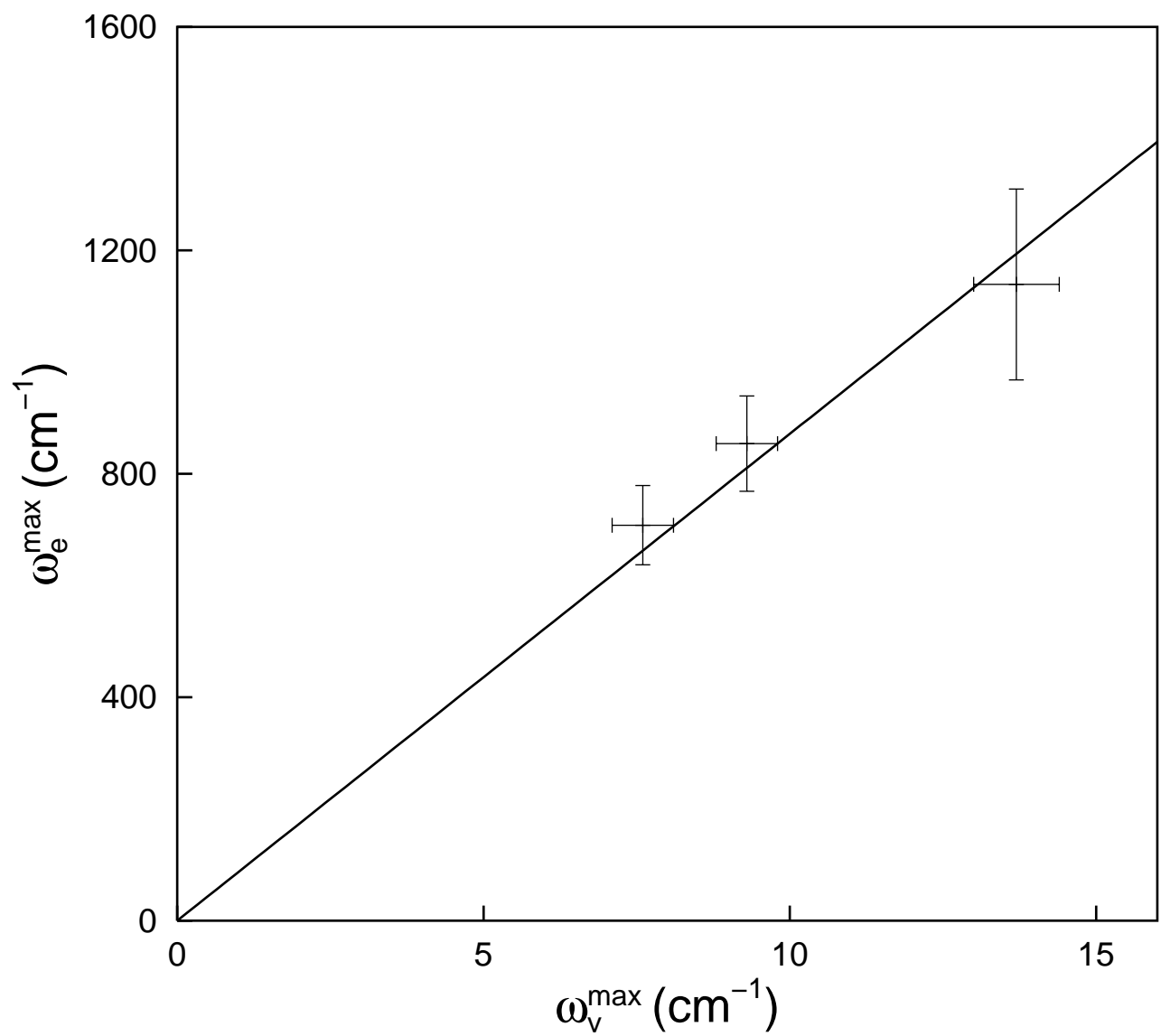
\end{document}